# Deep Learning-Enabled Invisible Electromagnetic Scattering Amplifier

Qike Xie,[1] Qin Liao,[1] Xiaofan Ji,[1] Yichao Liu,[1,*] and Fei Sun[1,*]

[1] *Key Lab of Advanced Transducers and Intelligent Control System, Ministry of Education and Shanxi Province, College of Physics and Optoelectronic Engineering, Taiyuan University of Technology, Taiyuan, 030024 China*
*\* liuyichao@tyut.edu.cn*
*\* sunfei@tyut.edu.cn*

**Abstract:** With the rapid development of micro-electro-mechanical systems, electrically small micro-targets, such as subwavelength micro unmanned aerial vehicles and bionic mosquito robots, exhibit ultra-low scattering cross section, which brings severe challenges to their effective detection. To address this problem, an Invisible Electromagnetic Scattering Amplifier (IESA) is designed by combining finite-element electromagnetic simulation with a forward lossless tandem neural network. The IESA realizes the dual-functional integration of intrinsic electromagnetic invisibility (near-zero scattering) for itself and significant scattering amplification for subwavelength targets entering its air sensing region. Electromagnetic simulations verify that the designed IESA can achieve a stable scattering amplification effect on subwavelength targets with a characteristic size of approximately $0.1\lambda_0$, regardless of their spatial positions or geometric shapes, with a maximum scattering cross section amplification factor of 8.58. The IESA breaks the technical bottleneck of the separate design of electromagnetic invisibility and scattering amplification functions. It shows potential for applications in the fields of radar detection, anti-terrorism security, micro-target monitoring, and adaptive electromagnetic sensing.

## 1. Introduction

With the rapid development of micro-robotic technology [1–4], emerging designs such as micro unmanned aerial vehicles (UAVs) and bionic mosquito robots have created an urgent demand for intelligent electromagnetic (EM) devices that integrate environmental concealment with active sensing capabilities. Such devices can maintain their inherent EM invisibility while significantly enhancing the EM scattering response of tiny, freely moving targets within the air sensing region, thereby achieving the strategic objective of "concealing itself while exposing the adversary." These EM devices hold broad application prospects in adaptive monitoring, security protection, and related scenarios.

To achieve the core requirement of "hiding itself while amplifying the target's scattering signal," previous research has often addressed the two functions of invisibility and scattering amplification separately. One research direction focuses on designing special cloaking structures to eliminate the scattering cross section (SCS) of the target object. For example, by employing transformation optics [5–7] or scattering cancellation theory [8], specially designed metamaterial shells can be created to either guide EM waves around the hidden object and restore their original propagation direction (light-bending cloaks) [9–14] or cancel out the scattering produced by the hidden object in response to external probing waves (scattering-cancellation cloaks) [15–18], thereby achieving the function of "invisibility" [19]. Another research direction aims to amplify the SCS of the target object by designing special shell structures, such as superscatterers based on spatial folding transformation [20,21], superscatterers designed using Mie scattering theory to break the single-channel scattering limit of objects [22,23] and amplification structures based on gain media or nonlinear effects [24]. Although both technologies have made significant progress in their respective fields of invisibility and scattering amplification, the core challenge at present lies in the fact that the synergistic integration of these two functions within a single device has not yet been realized.

With the advancement of artificial intelligence (AI) technology, new methods have emerged for designing novel optoelectronic devices [25–29]. While research groups have already utilized AI, such as deep learning for inverse design [30–32], to create invisibility cloaks [33,34], the academic community has not yet developed—through AI or other means—an effective structure capable of simultaneously achieving EM invisibility for itself and significantly amplifying the SCS of electrically small targets entering its air sensing region. To address this technological gap, this study proposes a forward lossless tandem neural network to design a novel functional device called the Invisible Electromagnetic Scattering Amplifier (IESA), whose core working mechanism is illustrated in Fig. 1. The designed IESA employs a multilayer cylindrical shell configuration. From the center outward, the structure includes: a central dielectric cylinder serving as a balanced medium, an air sensing region for micro-target detection, and four concentric dielectric shells. The permittivity and radii of the dielectric materials are optimized as variables. As shown in Fig. 1(a), when an electrically small micro-UAV is near a building but outside the IESA's air sensing region, the IESA itself maintains near-invisibility with the SCS close to zero, while the UAV's signal remains too weak for conventional detectors like radar or cameras to capture, leaving only background building visible. However, once the UAV enters the IESA's air sensing region (Fig. 1(b)), the IESA enhances the UAV's SCS so that conventional detection equipment can clearly identify it, thereby enabling precise detection of such minute invasive targets.

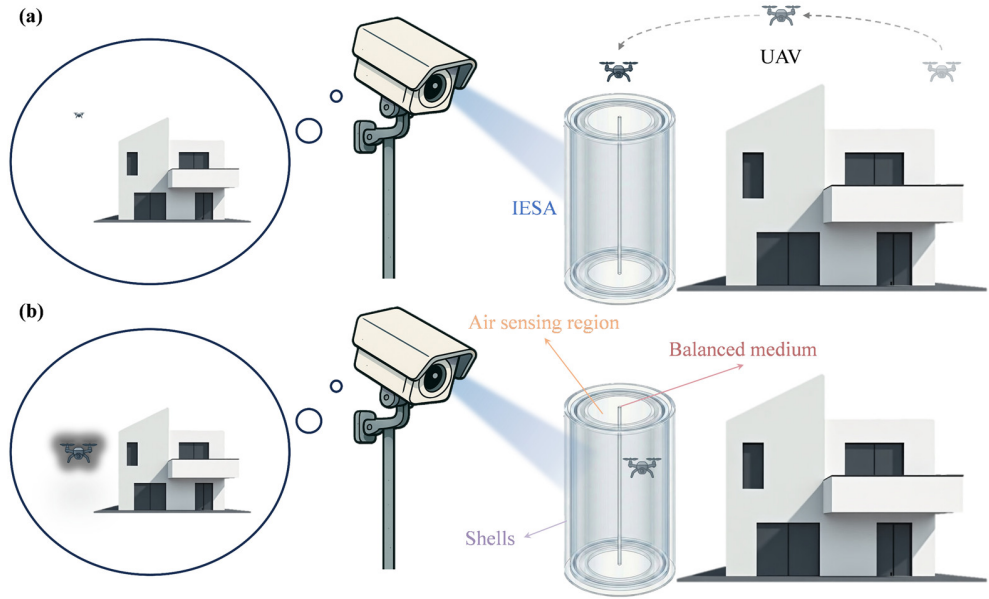

Fig. 1. Schematic diagram of the IESA operation around a building. (a) When an electrically small micro-UAV appears around the building but remains outside the IESA's air sensing region, the IESA stays invisible due to its negligible intrinsic EM scattering. Meanwhile, the micro-UAV, which has an ultra-low SCS, appears extremely small and is nearly undetectable by conventional detection systems, including radar and surveillance cameras. Consequently, only the background building is clearly visible in the surveillance field of view (see the left panel of Fig. 1(a)). (b) When the micro-UAV inadvertently enters the IESA's air sensing region via the IESA's upper opening, the IESA will significantly amplify the UAV's SCS. As a result, conventional detection systems can capture the micro-UAV, which exhibits enhanced scattering responses (see the left panel of Fig. 1(b)). The IESA consists of a balanced medium, an air sensing region, and four dielectric shell layers.

## 2. Method

**A. Pre-Training Settings for the IESA**

The basic IESA structure is shown in Fig. 2(a). The main body consists of a multilayer cylindrical structure with openings at both ends. These openings allow electrically small targets to enter the air sensing region, whose outer diameter is fixed at $r_0$. At the center is a dielectric cylinder with permittivity $\varepsilon_b$ and radius $r_b$, which serves as the balanced medium. The outer region is sequentially surrounded by four dielectric shells. The permittivity and outer radius of each layer are denoted as $\varepsilon_1$, $\varepsilon_2$, $\varepsilon_3$, $\varepsilon_4$ and $r_1$, $r_2$, $r_3$, $r_4$, respectively. The permittivity and radii of the dielectric materials are optimized as variables. Owing to the high geometric symmetry of the IESA model and to balance simulation accuracy with computational cost, a two-dimensional simulation framework is employed. A transverse magnetic (TM) plane wave is incident from the left, with a wavelength of $\lambda_0 = 60$ mm, which is twice $r_0$. The micro-target is modeled to have a characteristic size of approximately $0.1\lambda_0$ and is assumed to be a perfect electric conductor (PEC).

**B. Construction and Training Optimization of a Forward Lossless Tandem Neural Network**

Neural networks for inverse design aim to accurately map the highly nonlinear relationship between the structural parameters $S$ of the IESA and its EM response $R$. However, training is often constrained by the non-uniqueness of the mapping, in which different device structures may produce identical EM responses. A common strategy is to abandon constraints on the structural parameter space and introduce a physical constraint module to compute the EM response $R_{predict}$ of the predicted structure $S_{predict}$. For this purpose, tandem neural network architectures [35–37] are widely adopted in inverse design tasks. This architecture connects an inverse network in series with a pre-trained forward network (see Fig. 2(b)), which serves as the physical constraint module.

    However, the overall performance of conventional tandem frameworks is inherently constrained by the approximation capability of the physical constraint module. The fitting error inherent in the forward network limits the achievable upper bound of the inverse network accuracy. To overcome this issue, this work proposes a forward lossless tandem neural network that mitigates the accuracy degradation induced by the forward network. A two-stage training strategy is employed to improve the accuracy of the inverse network while controlling computational cost.

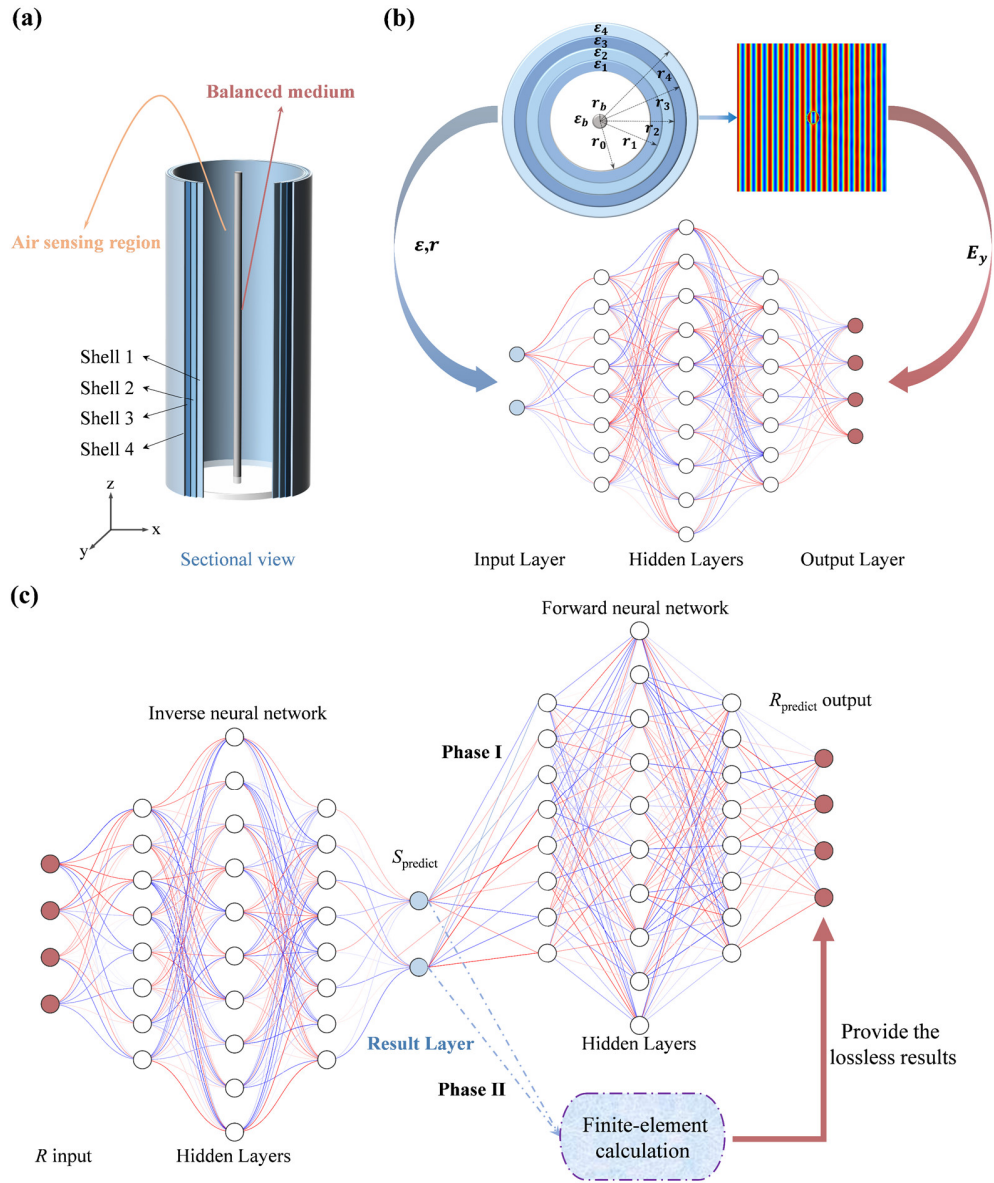

Fig. 2. Forward lossless tandem neural network architecture and workflow. (a) Three-dimensional schematic diagram of the IESA. (b) Forward neural network: The inputs are the structural parameters, including the radius $r$ and relative permittivity $\varepsilon$, and the output is the y-component of the electric field. After training, freeze the weights and biases. The network architecture consists of an input layer, four hidden layers containing 200–300–300–200 nodes, and an output layer. (c) In the first stage, the tandem architecture comprises an inverse network (including an input layer, four hidden layers with neuron counts of 300–400–400–400, and an output layer) cascaded with a pre-trained forward network. The inverse network takes the y-component of the electric field as input and outputs the predicted IESA structure. Building upon this framework, during the second training phase of the inverse network, finite-element simulations performed using COMSOL Multiphysics replace the forward network for computing the predicted response $R_{\text{predict}}$ from the predicted structure $S_{\text{predict}}$.

## C. Two-Stage Training Strategy

Phase I: Using finite-element EM simulation with COMSOL Multiphysics, construct a dataset containing 50,000 sets of structure-response data. EM responses are sampled at two characteristic wavelength-spaced distances, yielding a total of 60 discrete spectral points. The dataset is partitioned into training, validation, and test sets in an 8:1:1 ratio. The forward network is first trained to convergence using early stopping criteria. Subsequently, its weights and biases are frozen and embedded into the cascaded architecture. The forward network functions as a physical constraint module, while early stopping ensures stable convergence of the inverse network.

Phase II: To mitigate the inherent approximation errors of the forward network, a fine-tuning strategy based on a finite-element numerical solver is employed, as shown in Fig. 2(c). In this phase, the structural parameters predicted by the inverse network are directly fed into a finite-element EM solver. This solver serves as a lossless physical constraint module to obtain EM responses. Concurrently, the forward network no longer computes EM responses but instead provides approximate gradients for updating the inverse network via the chain rule. As shown in Figs. 3(a) and 3(b), the two-stage training strategy significantly enhances the sensitivity of the inverse network to extreme physical parameters and improves prediction accuracy.

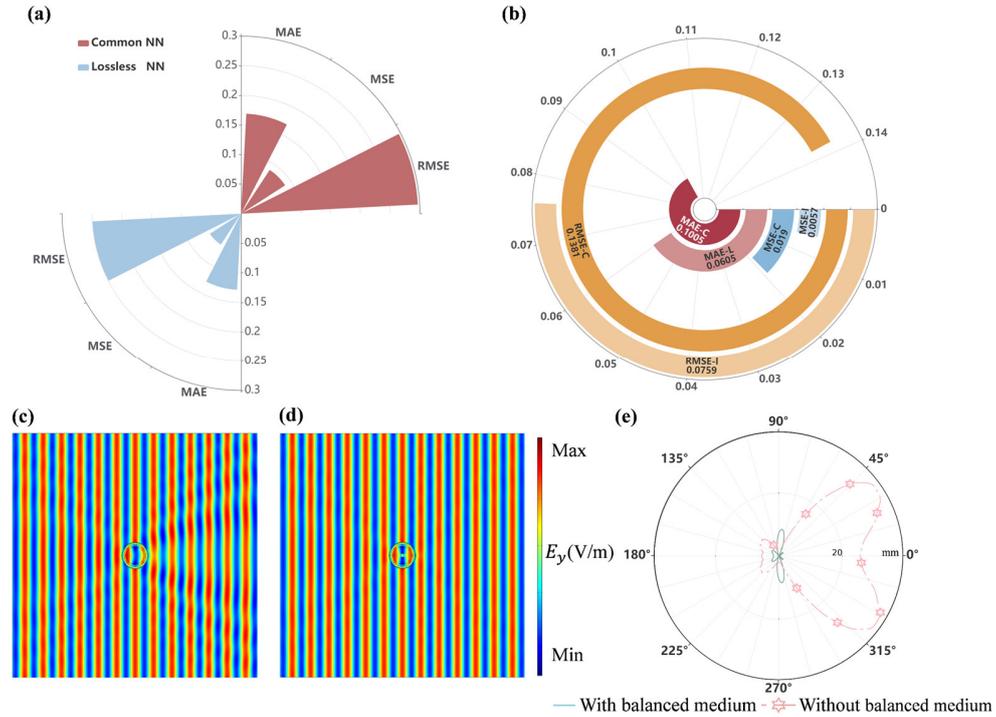

Fig. 3. Performance evaluation and balanced mechanism analysis of tandem neural networks under different physical-constraint modules. (a) Comparison of the test-set prediction performance between the forward lossless tandem neural network (Lossless NN), which uses COMSOL Multiphysics as the physical-constraint module during secondary training, and the conventional tandem neural network (Common NN) without secondary training. (b) Comparison of IESA structure prediction performance between the Common NN and the Lossless NN. (c) Near-field distribution of the final IESA without the balanced medium. (d) Near-field distribution of the final IESA with a balanced medium. IESA parameters: $r_1 = 36$ mm, $r_2 = 40.8$ mm, $r_3 = 43$ mm, $r_4 = 43.4$ mm; $\varepsilon_1 = 2.17$, $\varepsilon_2 = 25$, $\varepsilon_3 = 11$, $\varepsilon_4 = 9.2$ and $r_b = 6$ mm; $\varepsilon_b = 10.2$. (e) Comparison of the IESA's differential SCS with and without the balanced medium.

**D. The Balanced Medium**

In practical application scenarios, our training and simulation results show that satisfactory intrinsic invisibility can be achieved even without a central balanced medium and without imposing constraints on material parameters (e.g., permittivity) or geometric dimensions (e.g., radius) of the IESA cylindrical shell layers. However, considering manufacturing feasibility, the permittivity of commercially available materials is typically limited to the range of 1–25. In addition, each shell layer must be thicker than 0.4 mm to satisfy machining precision, structural stability, and manufacturability requirements.

Under the constraints, increasing the number of layers alone does not achieve the desired invisibility performance without a balanced medium, as shown in Fig. 3(c). Inspired by the bias term in neural networks, we introduce the concept of a balanced medium. By placing a balanced medium in the air sensing region, part of the scattered field is compensated, thereby significantly enhancing the IESA's invisibility performance, as shown in Fig. 3(d). Fig. 3(e) further demonstrates the improvement in invisibility performance provided by the balanced medium through a comparison of differential SCS.

In addition, the balanced medium provides the physical basis for scattering enhancement and improves robustness against structural deviations. When small foreign objects (such as UAVs) enter the air sensing region, they disrupt the original equilibrium state and trigger significant scattering enhancement. Even in invisible designs, deviations between fabricated and simulated geometric and material parameters are difficult to avoid. The balanced medium strategy substantially improves tolerance to such deviations. By flexibly adjusting the dimensions or permittivity of the balanced medium, manufacturing errors are effectively compensated.

## 3. Results

Based on finite-element EM numerical simulation combined with the forward lossless tandem neural network proposed above, the inverse design of the IESA structure is completed, and the final optimized structural parameters are obtained as shown in Fig. 3(d): $r_1 = 36$ mm, $r_2 = 40.8$ mm, $r_3 = 43$ mm, $r_4 = 43.4$ mm; $\varepsilon_1 = 2.17$, $\varepsilon_2 = 25$, $\varepsilon_3 = 11$, $\varepsilon_4 = 9.2$ and $r_b = 6$ mm; $\varepsilon_b = 10.2$. The final IESA achieves intrinsic EM invisibility for itself under TM-polarized EM waves with a wavelength of $\lambda_0 = 60$ mm, and can simultaneously realize significant scattering amplification for micro-targets entering the air sensing region. The following numerical simulations are conducted to verify the scattering enhancement effect of the IESA on subwavelength scatterers with different shapes and spatial positions within the air sensing region.

The simulation results shown in Figs. 4(a)–(c) demonstrate the scattering amplification effect of the IESA on an electrically small micro-UAV target at three distinct positions within the air sensing region. The target is modeled as a PEC with a characteristic size of approximately $0.1\lambda_0$. The differential SCS plots quantitatively characterize the scattering amplification performance, as detailed below. The standalone UAV exhibits an SCS of approximately $6.31\lambda_0^2$, $6.31\lambda_0^2$, and $4.87\lambda_0^2$ at the three positions (light-blue dashed lines with asterisks), respectively, all remaining at levels that are difficult to detect using conventional radar systems. In contrast, when the target couples with the IESA (light-green dotted lines with star-shaped markers), the SCS increases significantly to $47.93\lambda_0^2$, $47.94\lambda_0^2$, and $41.80\lambda_0^2$, respectively. The average amplification factor across the three positions is approximately 7.93, with Location 3 (corresponding to Fig. 4(c)) exhibiting the highest value of about 8.58.

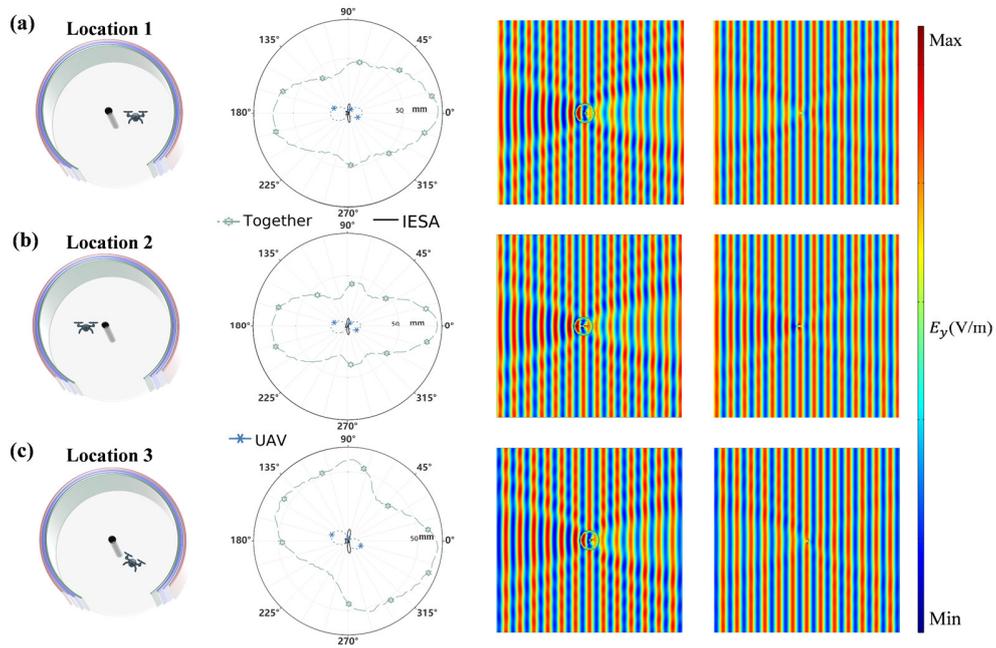

Fig. 4. Scattering amplification produced by the final IESA design for an electrically small micro-UAV at different positions in two-dimensional simulations. (a)–(c) Scattering amplification when the electrically small micro-UAV is placed at different locations within the IESA. For each subfigure, the columns from left to right are as follows: the first column is a schematic diagram showing the position of the micro-UAV in the IESA's air sensing region; the second column presents the differential SCS at different angles of the relevant structure; the third and fourth columns display the y-component of the electric field distribution when a TM-polarized plane wave is incident from the left on the micro-UAV with and without the IESA, respectively. In the differential SCS plots, black solid lines represent the standalone IESA response; light-blue dashed lines with asterisks denote the scattering response of the standalone electrically small micro-UAV; and light-green dotted lines with star-shaped markers indicate the coupled scattering response when the IESA and micro-UAV are present simultaneously.

Subsequently, we systematically investigate how the spatial position of electrically small micro-targets affects the scattering amplification effect. Figs. 5(a) and 5(b) illustrate cases in which the bionic mosquito is shifted horizontally to the left and right, respectively, relative to the equilibrium position in the balanced medium. Fig. 6(c) shows the response when the bionic mosquito rotates through 360° around the balanced medium, with the initial position located to the left of the balanced medium. Fig. 5(d) quantitatively shows how scattering amplification varies with the bionic mosquito's horizontal displacement, whereas Fig. 5(e) shows its dependence on the bionic mosquito rotation angle. Overall, the results indicate that the scattering amplification effect remains significant across different spatial orientations of the bionic mosquito.

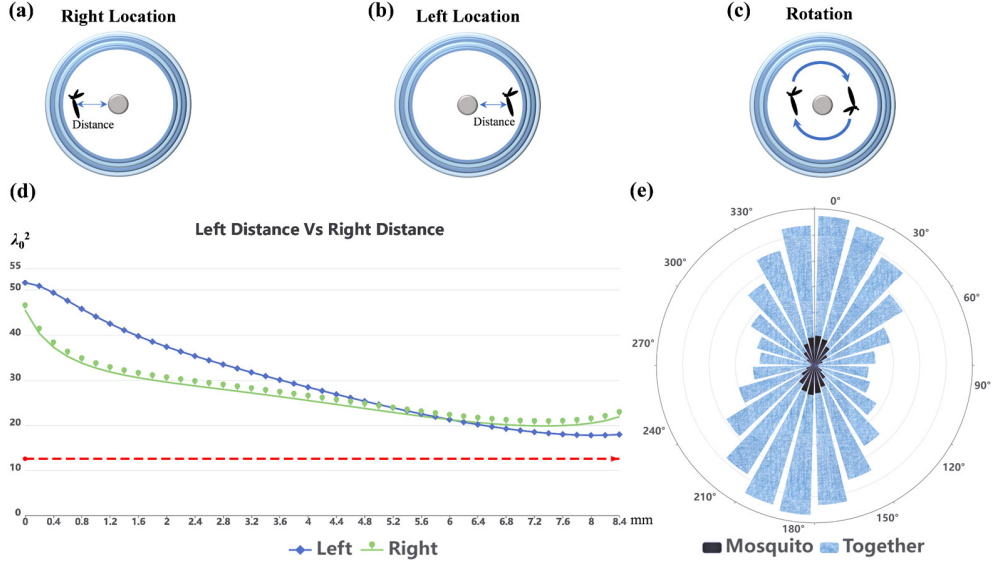

Fig. 5. Influence of target position and orientation on scattering amplification. (a)–(b) Bionic mosquito targets located at various positions within the IESA, where the position parameter is defined as the distance between the target and the geometric center of the balanced medium (mm). (c) The target is shown at different rotation angles (in degrees), starting from the left side of the balanced medium. (d) Position-dependent scattering-enhancement characteristics, comparing the IESA-induced scattering enhancement at different target positions. The red dashed line denotes the intrinsic scattering intensity of the bionic mosquito in free space, which serves as the baseline. The solid blue diamond line represents the scattering amplification effect at different positions on the left side of the balanced medium. The solid green line with triangle markers represents the scattering amplification effect at different positions on the right side of the balanced medium. (e) Angle-dependent scattering responses by analyzing the SCS of the isolated bionic mosquito and the IESA composite system at different orientation angles.

## 4. Conclusion

In conclusion, this work proposes an Invisible Electromagnetic Scattering Amplifier (IESA) based on the combination of finite-element EM simulation and the forward lossless tandem neural network, which innovatively realizes the synergistic integration of the device's own intrinsic EM invisibility and the scattering amplification of external electrically small targets. The optimized IESA maintains near-zero intrinsic EM scattering under TM-polarized EM wave excitation with a wavelength of 60 mm, and when electrically small micro-targets with a characteristic size of approximately $0.1\lambda_0$ enter its air sensing region, the device can stably enhance the scattering cross section of the targets regardless of the spatial position and geometric shape of the micro-targets, with a maximum amplification factor of 8.58. This research solves the challenge that the traditional EM devices cannot simultaneously realize self-invisibility and target scattering amplification, and the proposed design method and device structure provide a reliable technical solution for the high-sensitivity detection of micro-scale low-scattering targets. The IESA shows potential for applications in radar detection, security protection, micro-target monitoring and other related fields that require high-precision detection of micro low-scattering targets.

**Funding.**

National Natural Science Foundation of China (No. 12274317, No. 12374277), San Jin Talent Support Program—Shanxi Provincial Youth Top-notch Talent Project, the Natural Science Foundation of Shanxi Province (202303021211054), Shanxi Province Higher Education Institutions Young Faculty Research and Innovation Support Program (2025Q006), and Shanxi Province Postgraduate Practical Innovation Project.

**Disclosures.**

The authors declare no competing financial interests.

**Data availability.**

The data that support the findings of this study are available from the authors upon reasonable request.

**References**

1. H. Liu, J. Sun, Y. Zhang, Y. Zhang, C. Chen, P. Ci, and S. Dong, "Bioinspired, piezoelectrically-actuated deployable miniature robots," Mater. Sci. Eng.: R: Rep. **166**, 101054 (2025).
2. X. Wang, S. Jia, Y. Gao, C. Liu, Y. Wang, A. Liu, and W. Yang, "Optical-driven miniature robots: driving mechanism, applications and future trends," Lab Chip **25**, 4473–4507 (2025).
3. L. Yang, J. Jiang, F. Ji, Y. Li, K.-L. Yung, A. Ferreira, and L. Zhang, "Machine learning for micro- and nanorobots," Nat. Mach. Intell. **6**, 605–618 (2024).
4. M. Medany, L. Piglia, L. Achenbach, S. K. Mukkavilli, and D. Ahmed, "Model-based reinforcement learning for ultrasound-driven autonomous microrobots," Nat. Mach. Intell. **6**, 1076–1090 (2024).
5. J. B. Pendry, D. Schurig, and D. R. Smith, "Controlling electromagnetic fields," Science **312**, 1780–1782 (2006).
6. H. Chen, C. T. Chan, and P. Sheng, "Transformation optics and metamaterials," Nat. Mater. **9**, 387–396 (2010).
7. F. Sun, B. Zheng, H. Chen, W. Jiang, S. Guo, Y. Liu, Y. Ma, and S. He, "Transformation optics: from classic theory and applications to its new branches," Laser Photonics Rev. **11**, 1700034 (2017).
8. A. Alù and N. Engheta, "Achieving transparency with plasmonic and metamaterial coatings," Phys. Rev. E **72**, 16623 (2005).
9. Y. Huang, J. Zhang, Q. Yang, L. Meng, T. Yang, C.-W. Qiu, and Y. Luo, "Transformation-invariant laplacian metadevices robust to environmental variation," Adv. Mater. **37**, 2412929 (2025).
10. Y. Gao, Y. Luo, J. Zhang, Z. Huang, B. Zheng, H. Chen, and D. Ye, "Full-parameter omnidirectional transformation optical devices," Natl. Sci. Rev. **11**, nwad171 (2024).
11. Y. Liu, X. Ma, K. Chao, F. Sun, Z. Chen, J. Shan, H. Chen, G. Zhao, and S. Chen, "Simultaneously realizing thermal and electromagnetic cloaking by multi-physical null medium," Opto-Electron. Sci. **3**, 230027–230027 (2024).
12. B. Zheng, H. Lu, C. Qian, D. Ye, Y. Luo, and H. Chen, "Revealing the transformation invariance of full-parameter omnidirectional invisibility cloaks," Electromagn. Sci. **1**, 1–7 (2023).
13. Y. Liu, F. Sun, Y. Yang, Y. Hao, S. Liang, and Z. Wang, "A metamaterial-free omnidirectional invisibility cloak based on thrice transformations inside optic-null medium," Opt. Laser Technol. **157**, 108779 (2023).
14. D. Schurig, J. J. Mock, B. J. Justice, S. A. Cummer, J. B. Pendry, A. F. Starr, and D. R. Smith, "Metamaterial electromagnetic cloak at microwave frequencies," Science **314**, 977–980 (2006).
15. L. Lan, F. Sun, Y. Liu, C. K. Ong, and Y. Ma, "Experimentally demonstrated a unidirectional electromagnetic cloak designed by topology optimization," Appl. Phys. Lett. **103**, 121113 (2013).
16. G. Fujii and Y. Akimoto, "Electromagnetic-acoustic biphysical cloak designed through topology optimization," Opt. Express **30**, 6090 (2022).
17. J. Soric, Y. Ra'di, D. Farfan, and A. Alù, "Radio-transparent dipole antenna based on a metasurface cloak," Nat. Commun. **13**, 1114 (2022).
18. F. Bernal Arango, F. Alpeggiani, D. Conteduca, A. Opheij, A. Chen, M. I. Abdelrahman, T. F. Krauss, A. Alù, F. Monticone, and L. Kuipers, "Cloaked near-field probe for non-invasive near-field optical microscopy," Optica **9**, 684–692 (2022).
19. R. Sun, F. Sun, and Y. Liu, "Research progress and development trend of electromagnetic cloaking," Opto-Electron Eng. **51**, 5–35 (2024).
20. A. D. Yaghjian and S. Maci, "Alternative derivation of electromagnetic cloaks and concentrators," New J. Phys. **10**, 115022 (2008).
21. W. H. Wee and J. B. Pendry, "Shrinking optical devices," New J. Phys. **11**, 73033 (2009).
22. C. Qian, Y. Yang, Y. Hua, C. Wang, X. Lin, T. Cai, D. Ye, E. Li, I. Kaminer, and H. Chen, "Breaking the fundamental scattering limit with gain metasurfaces," Nat. Commun. **13**, 4383 (2022).
23. C. Wang, X. Chen, Z. Gong, R. Chen, H. Hu, H. Wang, Y. Yang, L. Tony, B. Zhang, H. Chen, and X. Lin, "Superscattering of light: fundamentals and applications," Rep. Prog. Phys. **87**, 126401 (2024).

24. J. Yi, E.-M. You, R. Hu, et al., "Surface-enhanced raman spectroscopy: a half-century historical perspective," Chem. Soc. Rev. **54**, 1453–1551 (2025).
25. X. Tian, R. Li, T. Peng, Y. Xue, J. Min, X. Li, C. Bai, and B. Yao, "Multi-prior physics-enhanced neural network enables pixel super-resolution and twin-image-free phase retrieval from single-shot hologram," Opto-Electron. Adv. **7**, 240060–240060 (2024).
26. Z. Yu, M. Li, Z. Xing, H. Gao, Z. Liu, S. Pu, H. Mao, H. Cai, Q. Ma, W. Ren, J. Zhu, and C. Zhang, "Genetic algorithm assisted meta-atom design for high-performance metasurface optics," Opto-Electron. Sci. **3**, 240016–240016 (2024).
27. W. Ma, Z. Liu, Z. A. Kudyshev, A. Boltasseva, W. Cai, and Y. Liu, "Deep learning for the design of photonic structures," Nat. Photonics **15**, 77–90 (2021).
28. Y. Saifullah, N. Wu, H. Wang, B. Zheng, C. Qian, and H. Chen, "Deep learning in metasurfaces: from automated design to adaptive metadevices," Adv. Photonics **7**, 0–16 (2025).
29. N. Wu, Y. Sun, J. Hu, C. Yang, Z. Bai, F. Wang, X. Cui, S. He, Y. Li, C. Zhang, K. Xu, J. Guan, S. Xiao, and Q. Song, "Intelligent nanophotonics: when machine learning sheds light," eLight **5**, 5–26 (2025).
30. Z. Yang, J. Duan, H. Xu, X. Li, S. Zhu, and H. Chen, "Multifrequency spherical cloak in microwave frequencies enabled by deep learning," Opt. Express **33**, 439–449 (2025).
31. J. Peurifoy, Y. Shen, L. Jing, Y. Yang, F. Cano-Renteria, B. G. DeLacy, J. D. Joannopoulos, M. Tegmark, and M. Soljačić, "Nanophotonic particle simulation and inverse design using artificial neural networks," Sci. Adv. **4**, 0–7 (2018).
32. Z. Zhen, C. Qian, Y. Jia, Z. Fan, R. Hao, T. Cai, B. Zheng, H. Chen, and E. Li, "Realizing transmitted metasurface cloak by a tandem neural network," Photonics Res. **9**, B229 (2021).
33. J. Zhao, P. Zhu, Z. Wen, F. Tang, B. Zheng, R. Zhu, H. Qian, C. Qian, H. Lu, and H. Chen, "Adaptive transparent cloaking tunnel enabled by meta-reinforcement-learning metasurfaces," PhotoniX **7**, 2–18 (2026).
34. H. Lu, J. Zhao, P. Zhu, W. Song, S. Zhu, R. Zhu, B. Zheng, and H. Chen, "Neural network-assisted metasurface design for broadband remote invisibility," Adv. Funct. Mater. **35**, 2506085 (2025).
35. K. Kojima, M. H. Tahersima, T. Koike-Akino, D. K. Jha, Y. Tang, Y. Wang, and K. Parsons, "Deep neural networks for inverse design of nanophotonic devices," J. Lightwave Technol. **39**, 1010–1019 (2021).
36. N. Wu, Y. Jia, C. Qian, and H. Chen, "Pushing the limits of metasurface cloak using global inverse design," Adv. Opt. Mater. **11**, 2202130 (2023).
37. D. Liu, Y. Tan, E. Khoram, and Z. Yu, "Training deep neural networks for the inverse design of nanophotonic structures," ACS Photonics **5**, 1365–1369 (2018).